\documentclass[runningheads]{llncs}
\usepackage{graphicx}
\usepackage{multirow}
\usepackage{booktabs}
\usepackage{makecell}
\usepackage{threeparttable}
\usepackage{makecell}

\begin{document}
\title{Adversarial Uni- and Multi-modal Stream Networks for Multimodal Image Registration}
% Unsupervised Multimodal Deformable Registration via GAN-based Image Translation and Dual-Stream Learning Network
% \thanks{Supported by organization x.}}
%
\titlerunning{Adversarial Uni- and Multi-modal Stream Networks}
%Specific Authors
\author{Zhe Xu\inst{1,2} \and
Jie Luo\inst{2,3} \and
Jiangpeng Yan\inst{1} \and Ritvik Pulya\inst{2} \and Xiu Li\inst{1} \and William Wells III\inst{2} \and Jayender Jagadeesan\inst{2} }
%index{Xu, Zhe}
%index{Luo, Jie}
%index{Yan, Jiangpeng}
%index{Pulya, Ritvik}
%index{Li, Xiu}
%index{Wells, William}
%index{Jayender, Jagadeesan}

% \inst{1,2}\orcidID{0000-1111-2222-3333} \and

\authorrunning{Z. Xu et al.}
\institute{Shenzhen International Graduate School, Tsinghua University, China 
\and Brigham and Women’s Hospital, Harvard Medical School, USA \\\email{jayender@bwh.harvard.edu}
\and Graduate School of Frontier Sciences, The University of Tokyo, Japan \\
}

\maketitle              % typeset the header of the contribution
\begin{abstract}
Deformable image registration between Computed Tomography (CT) images and Magnetic Resonance (MR) imaging is essential for many image-guided therapies. In this paper, we propose a novel translation-based unsupervised deformable image registration method. Distinct from other translation-based methods that attempt to convert the multimodal problem (e.g., CT-to-MR) into a unimodal problem (e.g., MR-to-MR) via image-to-image translation, our method leverages the deformation fields estimated from both: (i) the translated MR image and (ii) the original CT image in a dual-stream fashion, and automatically learns how to fuse them to achieve better registration performance. The multimodal registration network can be effectively trained by computationally efficient similarity metrics without any ground-truth deformation. Our method has been evaluated on two clinical datasets and demonstrates promising results compared to state-of-the-art traditional and learning-based methods. 
\end{abstract}
\section{Introduction}
Deformable multimodal image registration has become essential for many procedures in image-guided therapies, e.g., preoperative planning, intervention, and diagnosis. Due to substantial improvement in computational efficiency over traditional iterative registration approaches, learning-based registration approaches are becoming more prominent in time-intensive applications. 
\subsubsection{Related work.} Many learning-based registration approaches adopt fully supervised or semi-supervised strategies. Their networks are trained with ground-truth deformation fields or segmentation masks \cite{cao2017deformable,sedghi2019semi,hu2018weakly,hu2018label,liu2019multimodal}, and may struggle with limited or imperfect data labeling. A number of unsupervised registration approaches have been proposed to overcome this problem by training unlabeled data to minimize traditional similarity metrics, e.g., mean squared intensity differences \cite{VM2018,zhao2019unsupervised,kuang2019faim,hu2019dual,de2019deep,mahapatra2018joint}. However, the performances of these methods are inherently limited by the choice of similarity metrics. Given the limited selection of multimodal similarity metrics, unsupervised registration approaches may have difficulties outperforming traditional multimodal registration methods as they both essentially optimize the same cost functions. A recent trend for multimodal image registration takes advantage of the latent feature disentanglement \cite{qin2019unsupervised} and image-to-image translation \cite{cao2018deep,tanner2018generative,wei2019synthesis}. Specifically, translation-based approaches use Generative Adversarial Network (GAN) to translate images from one modality into the other modality, thus are able to convert the difficult multimodal registration into a simpler unimodal task. However, being a challenging topic by itself, image translation may inevitably produce artificial anatomical features that can further interfere with the registration process.  \\

In this work, we propose a novel translation-based fully unsupervised multimodal image registration approach. In the context of Computed Tomography (CT) image to Magnetic Resonance (MR) image registration, previous translation-based approaches would translate a CT image into an MR-like image (tMR), and use tMR-to-MR registration to estimate the final deformation field $\phi$. In our approach, the network estimates two deformation fields, namely $\phi_\mathrm{s}$ of tMR-to-MR and $\phi_\mathrm{o}$ of CT-to-MR, in a dual-stream fashion. The addition of the original $\phi_\mathrm{o}$ enables the network to implicitly regularize $\phi_\mathrm{s}$ to mitigate certain image translation problems, e.g., artificial features. The network further automatically learns how to fuse $\phi_\mathrm{s}$ and $\phi_\mathrm{o}$ towards achieving the best registration accuracy. 

Contributions and advantages of our work can be summarized as follows:
 \begin{enumerate}
  	\item Our method leverages the deformation fields estimated from the original multimodal stream and synthetic unimodal stream to overcome the shortcomings of translation-based registration;
    \item We improve the fidelity of organ boundaries in the translated MR by adding two extra constraints in the image-to-image translation model Cycle-GAN.
\end{enumerate}
We evaluate our method on two clinically acquired datasets. It outperforms state-of-the-art traditional, unsupervised and translation-based registration approaches.

\section{Methods}
In this work, we propose a general learning framework for robustly registering CT images to MR images in a fully unsupervised manner. 

\begin{figure}
\includegraphics[width=\textwidth]{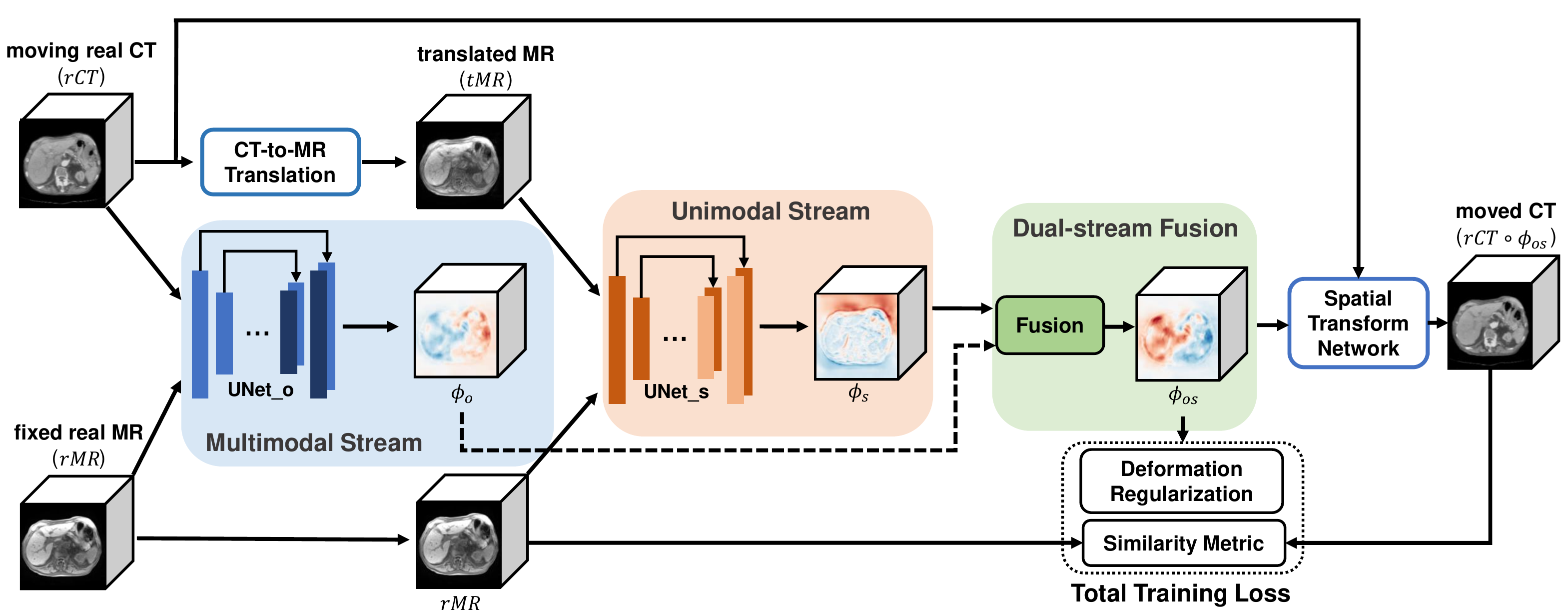}
\caption{ Illustration of the proposed method. The entire unsupervised network is mainly guided by the image similarity between ${\rm{rCT}} \circ {\phi _\mathrm{os}}$ and ${\rm{rMR}}$.} \label{fig1}
\end{figure}

First, given a moving CT image and a fixed MR image, our improved Cycle-GAN module translates the CT image into an MR-like image. Then, our dual-stream subnetworks, UNet\_o and UNet\_s, estimate two deformation fields ${\phi _\mathrm{o}}$ and ${\phi _\mathrm{s}}$ respectively, and the final deformation field is fused via a proposed fusion module. Finally, the moving CT image is warped via Spatial Transformation Network (STN)~\cite{STN}, while the entire registration network aims to maximize the similarity between the moved and the fixed images. The pipeline of our method is shown in Fig.~\ref{fig1}.

\subsection{Image-to-Image Translation with Unpaired Data}
The CT-to-MR translation step consists of an improved Cycle-GAN with additional structural and identical constraints. As a state-of-the-art image-to-image translation model, Cycle-GAN~\cite{zhu2017unpaired} can be trained without pairwise aligned CT and MR datasets of the same patient. Thus, Cycle-GAN is widely used in medical image translation \cite{yang2018unpaired,medGAN2019unsupervised,hiasa2018cross}.  

\begin{figure}
\centering
\includegraphics[width=4in]{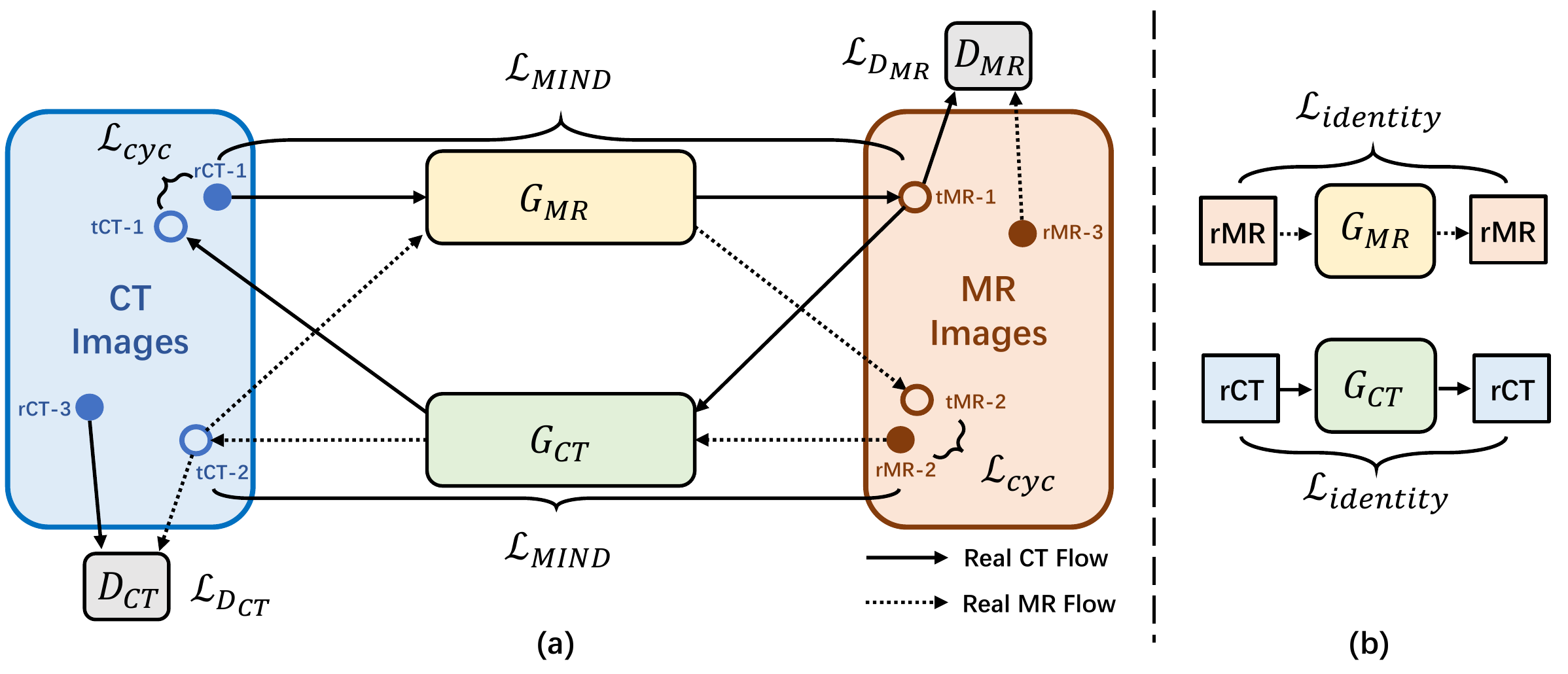}
\caption{ Schematic illustration of Cycle-GAN with strict constraints. (a) The workflow of the forward and backward translation; (b) The workflow of identity loss.} \label{fig2}
\end{figure}

Our Cycle-GAN model is illustrated in Fig.~\ref{fig2}. The model consists of two generators ${{\rm{G}}_{{\rm{MR}}}}$ and ${{\rm{G}}_{{\rm{CT}}}}$ , which can provide CT-to-MR and MR-to-CT translation respectively. Besides, it has two discriminators ${{\rm{D}}_{{\rm{CT}}}}$ and ${{\rm{D}}_{{\rm{MR}}}}$. ${{\rm{D}}_{{\rm{CT}}}}$ is used to distinguish between translated CT(tCT) and real CT(rCT), and ${{\rm{D}}_{{\rm{MR}}}}$ is for translated MR(tMR) and real MR(rMR). The training loss of original Cycle-GAN only adopts two types of items: adversarial loss given by two discriminators ($ {{\mathcal{L}}_{{D_{CT}}}}$ and ${{\mathcal{L}}_{{D_{MR}}}}$) and cycle-consistency loss ${{\mathcal{L}}_{{cyc}}}$ to prevent generators from generating images that are not related to the inputs (refer to~\cite{zhu2017unpaired} for details). 

\begin{figure}
\centering
\includegraphics[width=4.5in]{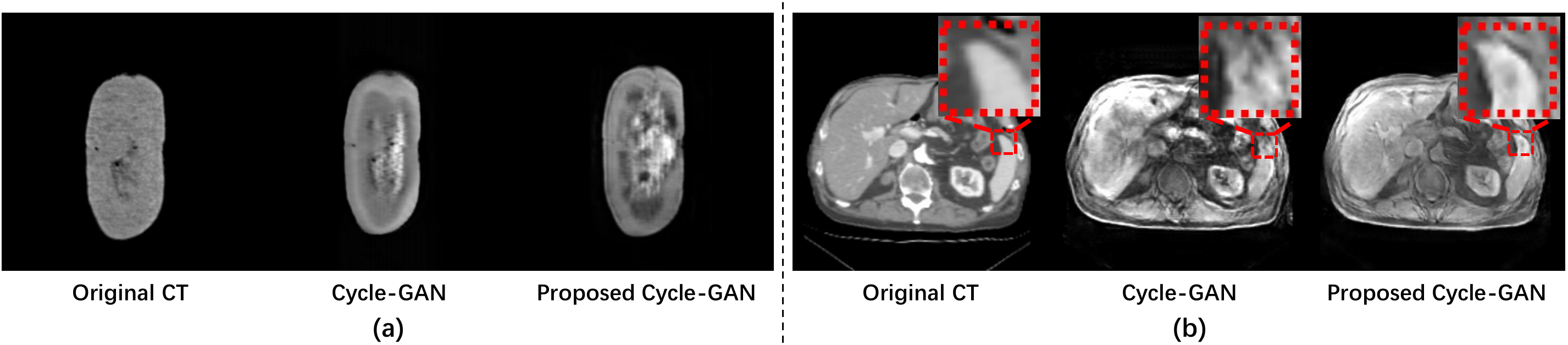}
\caption{CT-to-MR translation examples of original Cycle-GAN and proposed Cycle-GAN tested for (a) pig ex-vivo kidney dataset and (b) abdomen dataset.} \label{fig3}
\end{figure}

However, training a Cycle-GAN on medical images is difficult since the cycle-consistency loss is not enough to enforce structural similarity between translated images and real images (as shown in the red box in Fig.~\ref{fig3}(b)). Therefore, we introduce two additional losses, structure-consistency loss $ {\mathcal{L}}_{MIND}$ and identity loss ${\mathcal{L}}_{identity}$, to constrain the training of Cycle-GAN.

MIND (Modality Independent Neighbourhood Descriptor)~\cite{mind} is a feature that describes the local structure around each voxel. Thus, we minimize the difference in MIND between translated images $G_{CT} (I_{rMR})$ or $G_{MR} (I_{rCT})$ and real images $I_{rMR}$ or $I_{rCT}$ to enforce the structural similarity. We define ${\mathcal{L}}_{MIND}$ as follows:\\
\begin{equation}
\begin{array}{l}
{L_{MIND}}({G_{CT}},{G_{MR}}) = \frac{1}{{{N_{MR}}|R|}}\sum\nolimits_x {||M({G_{CT}}({I_{rMR}})) - M({I_{rMR}})|{|_1}} \\
{\kern 94.3pt}  + \frac{1}{{{N_{CT}}|R|}}\sum\nolimits_x {||M({G_{MR}}({I_{rCT}})) - M({I_{rCT}})|{|_1}} 
\end{array}
\end{equation}
where $M$ represents MIND features, $N_{MR}$ and $N_{CT}$ denote the number of voxels in $I_{rMR}$ and $I_{rCT}$, and $R$ is a non-local region around voxel $x$ .

The identity loss (as shown in Fig.~\ref{fig2}(b)) is included to prevent images already in the expected domain from being incorrectly translated to the other domain. We define it as:
\begin{equation}
\setlength\abovedisplayskip{4.5pt}
\setlength\belowdisplayskip{4.5pt}
\mathcal{L}_{identity}=\|G_{M R}(I_{M R})-I_{M R}\|_{1}+\| G_{C T}(I_{C T})-I_{C T} \|_{1}
\end{equation}

Finally, the total loss $\mathcal{L}$ of our proposed Cycle-GAN is defined as:
\begin{equation}
\setlength\abovedisplayskip{4.5pt}
\setlength\belowdisplayskip{4.5pt}
\mathcal{L}=\mathcal{L}_{D_{M R}}+\mathcal{L}_{D_{C T}}+\lambda_{c y c} \mathcal{L}_{c y c}+\lambda_{identity} \mathcal{L}_{identity}+\lambda_{M I N D} \mathcal{L}_{MIND}
\end{equation}
where $\lambda_{c y c}$,  $\lambda_{identity}$ and $\lambda_{MIND}$ denotes the relative importance of each term.
\subsection{Dual-stream Multimodal Image Registration Network}
As shown in Fig.~\ref{fig3}, although our improved Cycle-GAN can better translate CT images into MR-like images, the CT-to-MR translation is still challenging for translating “simple” CT images to “complex” MR images. Most image-to-image translation methods will inevitably generate unrealistic soft-tissue details, resulting in some mismatch problems. Therefore, the registration methods that simply convert multimodal to unimodal registration via image translation algorithm are not reliable.

In order to address this problem, we propose a dual-stream network to fully use the information of the moving, fixed and translated images as shown in Fig.~\ref{fig1}. In particular, we can use effective similarity metrics to train our multimodal registration model without any ground-truth deformation.

\subsubsection{Network Details.} As shown in Fig.~\ref{fig1},  our dual-stream network is comprised of four parts: multimodal stream subnetwork, unimodal stream subnetwork, deformation field fusion, and Spatial Transformation Network.

In \textbf{\textsl{Multimodal Stream}} subnetwork, original CT(rCT) and MR(rMR) are represented as the moving and fixed images, which allows the model to propagate original information to counteract mismatch problems in translated MR(tMR).

Through image translation, we obtain the translated MR(tMR) with similar appearance to the fixed MR(rMR). Then, in \textbf{\textsl{Unimodal Stream}}, tMR and rMR are used as moving and fixed images respectively. This stream can effectively propagate more texture information, and constrain the final deformation field to suppress unrealistic voxel drifts from the multimodal stream.

During the network training, the two streams constrain each other, while they are also cooperating to optimize the entire network. Thus, our novel dual-stream design allows us to benefit from both original image information and homogeneous structural information in the translated images. 

\begin{figure}
\centering
\includegraphics[width=4in]{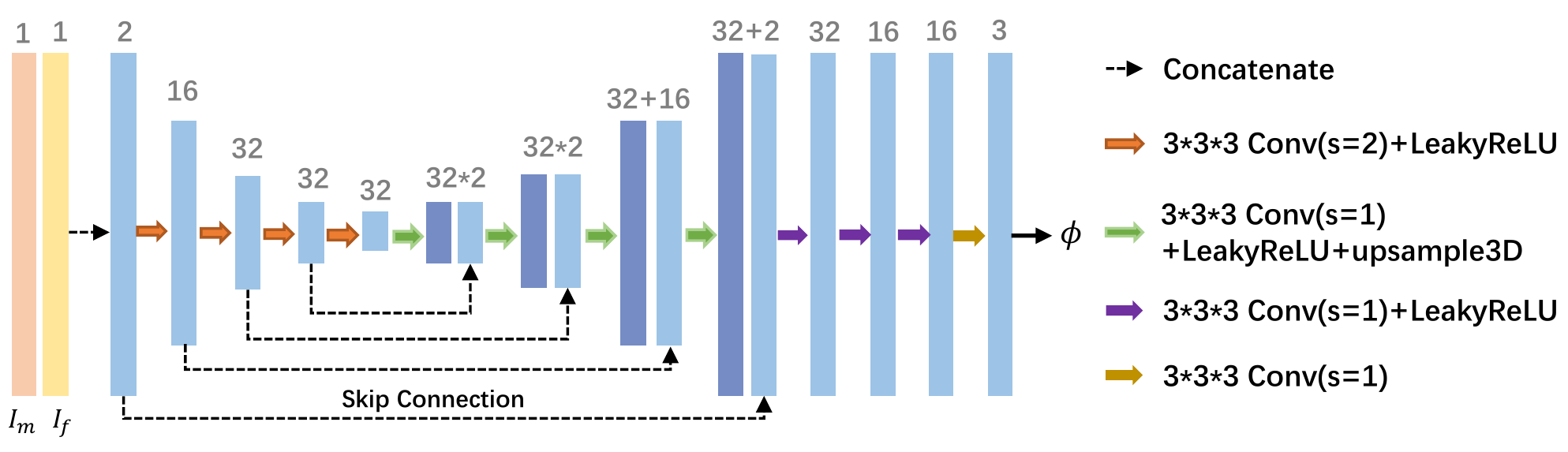} % or width=4.3in
\caption{Detailed architecture of UNet-based subnetwork. The encoder uses convolution with stride of 2 to reduce spatial resolution, while the decoder uses 3D upsampling layers to restore the spatial resolution.} \label{fig4}
\end{figure}

Specifically, UNet\_o and UNet\_s adopt the same UNet architecture used in VoxelMorph~\cite{VM2018} (shown in Fig.~\ref{fig4}) . The only difference is that UNet\_o is with \textsl{multimodal inputs} but UNet\_s is with \textsl{unimodal inputs}. Each UNet takes a single 2-channel 3D image formed by concatenating $I_m$ and $I_f$ as input, and outputs a volume of deformation field with 3 channels.

After Uni- and Multi-model Stream networks, we obtain two deformation fields, ${\phi_\mathrm{o}}$ (for rCT and rMR) and ${\phi _\mathrm{s}}$ (for tMR and rMR). We stack ${\phi _\mathrm{o}}$ and ${\phi _\mathrm{s}}$, and apply a 3D convolution with size of $3\times3\times3$ to estimate the final deformation field  ${\phi _\mathrm{os}}$, which is a 3D volume with the same shape of ${\phi _\mathrm{o}}$ and ${\phi _\mathrm{s}}$. 

To evaluate the dissimilarity between moved and fixed images, we integrate spatial transformation network (STN)~\cite{STN} to warp the moving image using ${\phi _\mathrm{os}}$. The loss function consists of two components as shown in Eq. (4).
\begin{equation}
\setlength\abovedisplayskip{5pt}
\setlength\belowdisplayskip{5pt}
\mathcal{L}_ {total}(I_{r M R}, I_{r C T}, \phi_\mathrm{os})=\mathcal{L}_{sim}(I_{r M R}, I_{r C T} \circ \phi_\mathrm{os})+\lambda \mathcal{L}_{smooth}(\phi_\mathrm{os})
\end{equation}
where $\lambda$ is a regularization weight. The first loss $\mathcal{L}_{sim}$ is similarity loss, which is to penalize the differences in appearance between fixed and moved images. Here we adopt SSIM~\cite{SSIM} for experiments. Suggested by~\cite{VM2018}, deformation regularization $\mathcal{L}_{smooth}$  adopts a L2-norm of the gradients of the final deformation field $\phi_\mathrm{os}$. 

\section{Experiments and Results}
\subsubsection{Dataset and Preprocessing.}
We focus on the application of abdominal CT-to-MR registration.We evaluated our method on two proprietary datasets since there is no designated public repository.\\

\noindent1) \textsl{Pig Ex-vivo Kidney CT-MR Dataset}. This dataset contains 18 pairs of CT and MRI kidney scans from pigs. All kidneys are manually segmented by experts. After preprocessing the data, e.g., resampling and affine spatial normalization, we cropped the data to $144\times80\times256$ with 1mm isotropic voxels and arbitrarily divided it into two groups for training (15 cases) and testing (3 cases). \\
\noindent2) \textsl{Abdomen (ABD) CT-MR Dataset}. This 50-patient dataset of CT-MR scans was collected from a local hospital and annotated with anatomical landmarks. All data were preprocessed into $176\times176\times128$ with the same resolution (${1mm}^{3}$) and were randomly divided into two groups for training (45 cases) and testing (5 cases).

\subsubsection{Implementation.}
We trained our model using the following settings: (1) The Cycle-GAN for CT-MR translation network is based on the existing implementation \cite{CycleGAN-code} with changes as discussed in Section 2.1. (2) The Uni- and Multi-modal stream registration networks were implemented using Keras with the Tensorflow backend and trained on an NVIDIA Titan X (Pascal) GPU.

\subsection{Results for CT-to-MR Translation}
We extracted 1792 and 5248 slices from the transverse planes of the Pig kidney and ABD dataset respectively to train the image translation network. Parameters $\lambda_{c y c}$, $\lambda_{identity}$ and $\lambda_{MIND}$ were set to 10, 5, and 5 for training. 

% \usepackage{multirow}
\begin{table}[]
\caption{Quantitative results for image translation.}\label{tab1}
\centering
\begin{tabular}{p{1.2cm}<{\centering}|c|p{1.3cm}<{\centering}|p{1.3cm}<{\centering}|p{1.2cm}<{\centering}|c|p{1.3cm}<{\centering}|p{1.3cm}<{\centering}}
\Xhline{1pt}
\hline
\multirow{3}{*}{\textbf{\begin{tabular}[c]{@{}c@{}}Pig\\ Kidney\end{tabular}}} & \textbf{Method} & \textbf{PSNR} & \textbf{SSIM} & \multirow{3}{*}{\textbf{ABD}} & \textbf{Method} & \textbf{PSNR} & \textbf{SSIM} \\ \cline{2-4} \cline{6-8} 
                                                                              & Cycle-GAN       &    32.07           &          0.9025     &                               & Cycle-GAN       & 22.95         & 0.7367        \\ \cline{2-4} \cline{6-8} 
                                                                               & Ours            &        \textbf{32.74}       &   \textbf{0.9532}            &                               & Ours            & \textbf{23.55}         & \textbf{0.7455}        \\ \Xhline{1pt}
\end{tabular}
\end{table}

Since our registration method is for 3D volumes, we apply the pre-trained CT-to-MR generator to translate moving CT images into MR-like images slice-by-slice and concatenate 2D slices into 3D volumes. The qualitative results are visualized in Fig.~\ref{fig3}. In addition, to quantitatively evaluate the translation performance, we apply our registration method to obtain aligned CT-MR pairs and utilize SSIM~\cite{SSIM} and PSNR~\cite{PSNR} to judge the quality of translated MR (shown in Table~\ref{tab1}). In our experiment, our method predicts better MR-like images on both datasets.

\subsection{Registration Results}
Affine registration is used as the baseline method. For traditional method, only mutual information (MI) based \textbf{SyN}~\cite{Avants2008SymmetricDI} is compared since it is the only metric (available in ANTs~\cite{Avants2011ARE}) for multimodal registration. In addition to SyN, we implemented the following learning-based methods: 1) \textbf{VM\_MIND} and \textbf{VM\_SSIM} which extends VoxelMorph with similarity metrics MIND~\cite{mind} and SSIM~\cite{SSIM}. 2) \textbf{M2U} which is a typical translation-based registration method. It generates tMR from CT and converts the multimodal problem to tMR-to-MR registration. It's noteworthy that the parameters of all methods are optimized to the best results on both datasets.

\begin{figure}
\includegraphics[width=\textwidth]{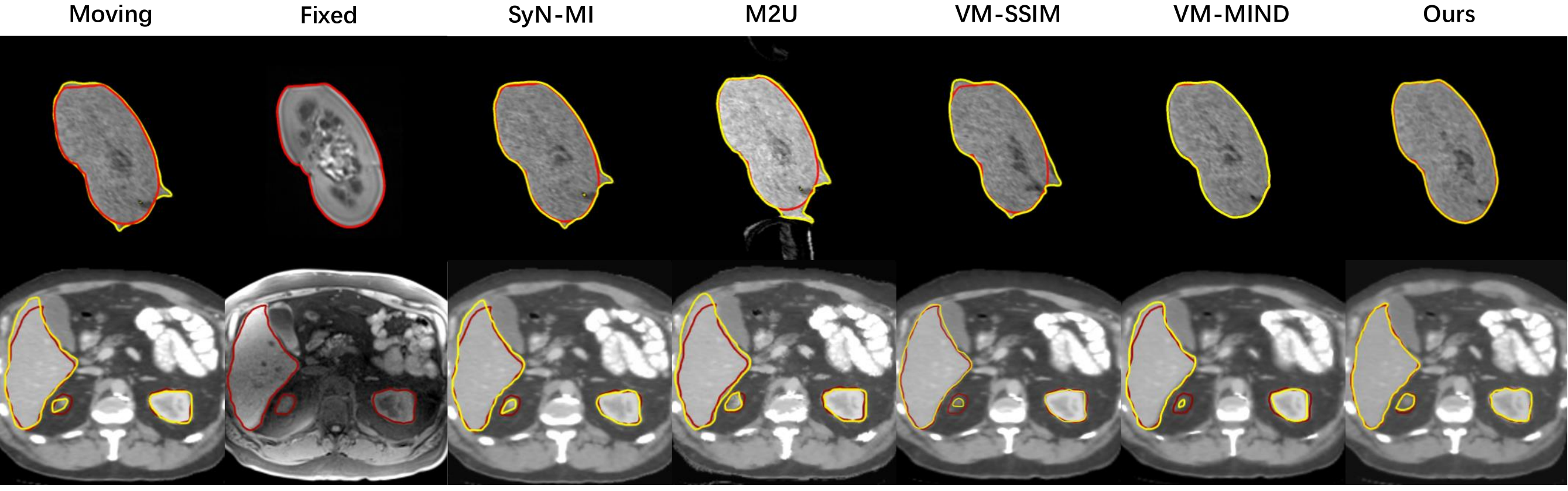}
\caption{Visualization results of our model compared to other methods. Upper: Pig Kidney. Bottom: Abdomen(ABD). The red contours represent the ground truth organ boundary while the yellow contours are the warped contours of segmentation masks.} \label{fig5}
\end{figure}

Two examples of the registration results are visualized in Fig.~\ref{fig5}, where the red and yellow contours represent the ground truth and registered organ boundaries respectively. As shown in Fig.~\ref{fig5}, the organ boundaries aligned by the traditional SyN method have a considerable amount of disagreement. Among all learning-based methods, our method has the most visually appealing boundary alignment for both cases. \textbf{VM\_SSIM} performed significantly worse for the kidney. \textbf{VM\_MIND} achieved accurate registration for the kidney, but its result for the ABD case is significantly worse. Meanwhile, \textbf{M2U} suffers from artificial features in the image translation, which leads to an inaccurate registration result.

The quantitative results are presented in Table~\ref{tab2}. We compare different methods by the Dice score~\cite{dice1945measures} and target registration error (TRE)~\cite{TRE}. We also provide the average run-time for each method. As shown in Table~\ref{tab2}, our method consistently outperformed other methods and was able to register a pair of images in less than 2 seconds (when using GPU).

% \usepackage{multirow}
\begin{table}[]\scriptsize
\caption{Quantitative results for Pig Kidney Dataset and Abdomen (ABD) Dataset.}\label{tab2}
\centering
\begin{tabular}{c|c|p{1.0cm}<{\centering}|p{1.3cm}<{\centering}|p{1.2cm}<{\centering}|p{1.2cm}<{\centering}|c|c|p{1.3cm}<{\centering}}
% \toprule
% \hline
\Xhline{1pt}
\textbf{Metric}                                                                     & \multicolumn{2}{c|}{\textbf{Organ}} & \textbf{Affine} & \textbf{SyN} & \textbf{M2U} & \textbf{VM\_SSIM} & \textbf{VM\_MIND}& \textbf{Ours}  \\ \hline
\multirow{4}{*}{\textbf{\begin{tabular}[c]{@{}c@{}}Dice\\ (\%)\end{tabular}}}       & Pig                     & kidney    & 89.53           & 89.87            &  90.21  & 93.75& 96.48 &\textbf{98.57} \\ \cline{2-9} 
                                                                                    & \multirow{3}{*}{ABD}    & kidney    &     80.03      &      82.36            &    78.96                     &  82.21 &84.58   &\textbf{85.66}           \\ \cline{3-9} 
                                                                                    &                         & spleen    &     79.58			            &        80.38          &      77.76                   &   81.79 & 83.11  &  \textbf{87.01}           \\ \cline{3-9} 
                                                                                    &                         & liver     &     78.74			           &     79.13             &      78.83                   &   82.05& 81.98   &  \textbf{83.34}           \\ \hline
\multirow{2}{*}{\textbf{\begin{tabular}[c]{@{}c@{}}TRE\\(mm)\end{tabular}}}        & \multirow{2}{*}{ABD}                             & spleen    &    4.16             &      4.20            &        3.76                 &    3.58 &  3.65  & \textbf{2.47}         \\ \cline{3-9} 
                                                                                    &                         & liver     &     6.55            &        5.61          &       5.91                  &    4.72 & 4.87& \textbf{3.64}            \\ \hline
\multirow{2}{*}{\textbf{\begin{tabular}[c]{@{}c@{}}Time(s)\\ GPU/CPU\end{tabular}}} & \multicolumn{2}{c|}{Pig}            & -/103           & -/121            &      1.08/20       & 1.06/20 & 1.07/20  &  1.12/21        \\ \cline{2-9} 
                                                                                    & \multicolumn{2}{c|}{ABD}            & -/108           & -/137            &         1.23/24   &  1.22/22 & 1.21/23 &  1.27/24      \\ \Xhline{1pt}
\end{tabular}
\end{table}
\subsection{The effect of each deformation field}
\begin{figure}
\includegraphics[width=\textwidth]{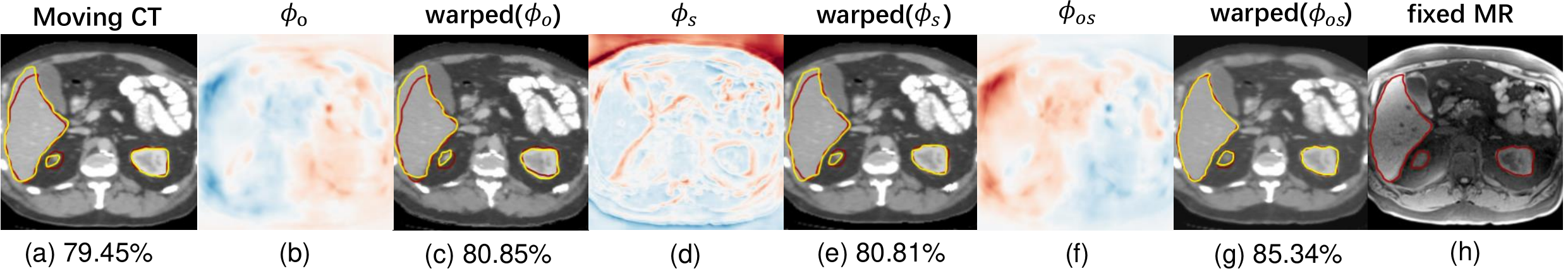}
\caption{Visualizations of the deformation field fusion. (a) moving image; (h) fixed image; (b/d/f) deformation fields; (c/e/g) images warped by (b/d/f), corresponding average Dice scores (\%) of all organs are calculated. The contours in red represent ground truth, while yellow shows the warped segmentation mask.} \label{fig6}
\end{figure}

In order to validate the effectiveness of the deformation field fusion, we compare $\phi_\mathrm{s}$, $\phi_\mathrm{o}$ and $\phi_\mathrm{os}$ together with warped images (shown in Fig.~\ref{fig6}). The qualitative result shows that $\phi_\mathrm{s}$ from the unimodal stream alleviates the voxel drift effect from the multimodal stream. While $\phi_\mathrm{o}$ from the multimodal stream uses the original image textures to maintain the fidelity and reduce artificial features for the generated tMR image. The fused deformation field $\phi_\mathrm{os}$ produces better alignment than both streams alone, which demonstrates the effectiveness of the joint learning step. 

.

\section{Conclusion}
We proposed a fully unsupervised uni- and multi-modal stream network for CT-to-MR registration. Our method leverages both CT-translated-MR and original CT images towards achieving the best registration result. Besides, the registration network can be effectively trained by computationally efficient similarity metrics without any ground-truth deformation. We evaluated the method on two clinical datasets, and it outperformed state-of-the-art methods in terms of accuracy and efficiency. 

\subsubsection{Acknowledgement}
This project was supported by the National Institutes of Health (Grant No. R01EB025964, R01DK119269, and P41EB015898) and the Overseas Cooperation Research Fund of Tsinghua Shenzhen International Graduate School (Grant No. HW2018008).

\bibliographystyle{splncs04.bst}
\bibliography{refs.bib}
\end{document}